\documentclass[reprint,superscriptaddress,amsmath,amsfonts,amssymb,aps,prl,showkeys]{revtex4-2}
\usepackage{hyperref}
\usepackage{graphicx}
\hypersetup{
  breaklinks = {true},
  citecolor = {blue},
  colorlinks = {true},
  linkcolor = {red},
}

\begin{document}

\title{Emerging Spin-Orbit Torques in Low Dimensional Dirac Materials}

\author{Joaqu{\'i}n Medina Due{\~n}as}
\affiliation{ICN2 --- Catalan Institute of Nanoscience and Nanotechnology, CSIC and BIST, Campus UAB, Bellaterra, 08193 Barcelona, Spain}
\affiliation{Department of Physics, Universitat Aut{\`o}noma de Barcelona (UAB), Campus UAB, Bellaterra, 08193 Barcelona, Spain}
 \author{Jos{\'e} H. Garc{\'i}a}
\affiliation{ICN2 --- Catalan Institute of Nanoscience and Nanotechnology, CSIC and BIST, Campus UAB, Bellaterra, 08193 Barcelona, Spain}
\author{Stephan Roche}
\affiliation{ICN2 --- Catalan Institute of Nanoscience and Nanotechnology, CSIC and BIST,
Campus UAB, Bellaterra, 08193 Barcelona, Spain}
\affiliation{ICREA --- Instituci\'o Catalana de Recerca i Estudis Avan\c{c}ats, 08010 Barcelona, Spain}

\date{\today}

\begin{abstract}
We report a theoretical description of novel spin-orbit torque components emerging in two-dimensional Dirac materials with broken inversion symmetry. In contrast to usual metallic interfaces where field-like and damping-like torque components are competing, we find that an intrinsic damping-like torque which derives from all Fermi-sea electrons can be simultaneously enhanced along with the field-like component. Additionally, hitherto overlooked torque components unique to Dirac materials, emerge from the coupling between spin and pseudospin degrees of freedom. These torques are found to be resilient to disorder and could enhance the magnetic switching performance of nearby magnets.
\end{abstract}

\maketitle

Spin-orbit torque (SOT) nonvolatile magnetic memories represent an emerging technology that leverages the intrinsic spin-orbit coupling (SOC) within metals to convert charge current into a spin source, denoted as $\boldsymbol{S}$, further harnessed to manipulate a magnetic state \cite{ando_electric_2008,miron_current-driven_2010}. When oriented along unit vector $\hat{\boldsymbol{m}}$, this magnetic state is subjected to a torque $\boldsymbol{T}$ proportional to $\hat{\boldsymbol{m}} \times \boldsymbol{S}$ driving its magnetization dynamics and eventually achieving magnetization reversal \cite{manchon_current-induced_2019}. This mechanism offers new possibilities regarding energy efficiency and miniaturization of devices, surpassing traditional multi-ferromagnet setups \cite{StephanNature2022}.

SOT is typically broken down into two main contributions: the field-like (FL) and damping-like (DL) torques. The FL torque induces precession of the magnetization, while the DL torque aligns it along an effective spin-orbit field \cite{haney_current_2013,ramaswamy_recent_2018,manchon_current-induced_2019}. For efficient magnetic switching, both strong FL and DL torques are necessary; however, in standard metal/magnet interfaces they have competing origins. The FL torque usually stems from the Rashba-Edelstein effect enabled by the reduced symmetry at the interface, whereas the DL torque arises from the injection of angular momentum from the metallic bulk states towards the magnet via the spin Hall effect \cite{kim_layer_2013, ramaswamy_hf_2016}. In this context, Van der Waals heterostructures offer a novelty to diversify and simultaneously enhance the efficiency of both torque components. Indeed, interfacial effects dominate the torque physics in such heterostructures, so the proper design of crystal symmetries can enable novel torque responses with tailor-made properties for driving the magnetization dynamics of nearby magnetic materials \cite{kurebayashi_magnetism_2022}. Surprisingly, despite the suppression of the perpendicular spin Hall effect, DL torque has been reported and tentatively explained in terms of Berry curvature effects \cite{kurebayashi_antidamping_2014,ghosh_spin-orbit_2018}. Additionally, skew-scattering has been theoretically proposed as a potential source of extrinsic interfacial DL torques, but the lack of microscopic information concerning interface quality does not facilitate a convincing explanation \cite{sousa_skew-scattering-induced_2020, zollner_scattering-induced_2020}. 

Dirac materials present favorable conditions for efficient FL torque due to their unique spin-momentum locking \cite{pesin_spintronics_2012}. Specifically, transport conveyed by a single spin-helical band 
represents an optimal regime for the Rashba-Edelstein effect. Such phenomenon has been manifested by the edge states of 3D topological insulators \cite{mellnik_spin-transfer_2014} and in graphene/transition metal dichalcogenide (TMD) bilayers \cite{benitez_tunable_2020,hoque_all-electrical_2021,offidani_optimal_2017}. Although magnetic exchange is detrimental to the Rashba-Edelstein effect, new torque components beyond the FL contribution can be created by reducing the system's symmetry \cite{garate_inverse_2010}. For instance, recent measurements with $\text{WTe}_2\text{/Py}$ \cite{macneill_control_2017} and $\text{CuPt/CoPt}$ \cite{liu_symmetry-dependent_2021} interfaces have reported the emergence of unconventional torques, leading to the strongly desired {\it field-free magnetization switching}.

In this Letter, we develop a general theory for SOT mechanisms in 2D-based materials, based on symmetry analysis, semi-classical modeling, and quantum simulations in realistic electronic models. We first employ group theory to determine the minimal set of torque contributions common to all 2D systems, beyond the standard FL contribution. We then use Boltzmann transport theory to elucidate the nature of the emerging torques. We finally quantify the SOT response for a few illustrative cases using Kubo quantum transport simulations. We reveal that the origin of a DL torque in 2D-based systems derives from the spin-momentum locking of accelerating electronic states and manifests through the entire Fermi sea. Furthermore, we find that Dirac materials present hitherto overlooked torque contributions which can induce non-trivial magnetization dynamics.

\begin{figure}
    \centering
    \includegraphics[width=\columnwidth]{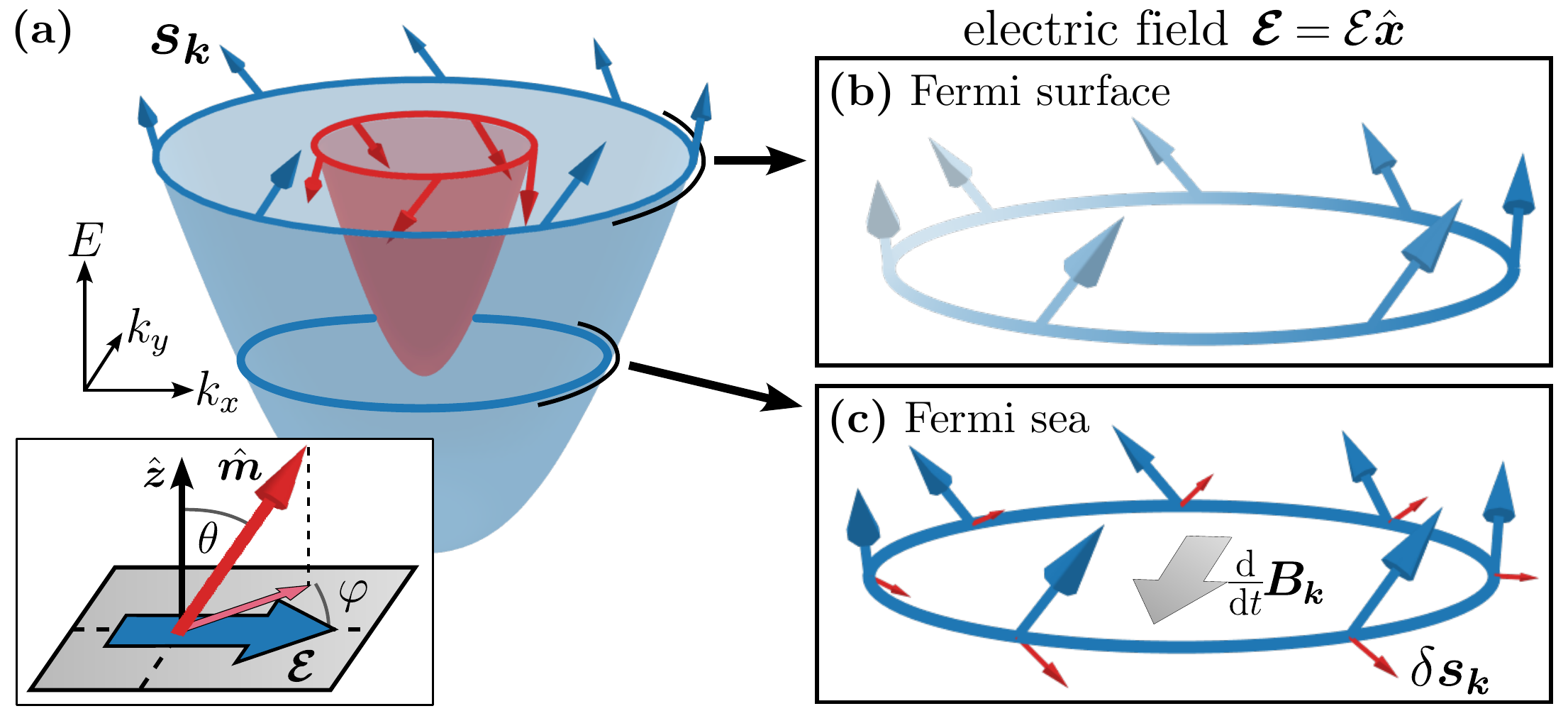}
    \caption{Physical origin of the FL and DL torques. \textbf{(a)} Spin texture of a 2DEG with Rashba SOC and out-of-plane magnetization ($\theta=0$). Inset: scheme of the magnetization orientation. When applying an electric field $\boldsymbol{\mathcal{E}}$: \textbf{(b)} the Fermi surface drifts overpopulating states with $\boldsymbol{k} \cdot \boldsymbol{\mathcal{E}} > 0$, generating a non-equilibrium spin density along $\boldsymbol{H}_\text{SOC} = \hat{\boldsymbol{z}} \times \boldsymbol{\mathcal{E}}$. \textbf{(c)} The acceleration of the carriers generates a forcing $\sim \frac{\text{d}}{\text{d}t} \boldsymbol{B}_{\boldsymbol{k}}$ (gray). Interaction between the equilibrium spins (blue) and the forcing originates a current-induced spin texture (red), responsible for the DL torque.}
    \label{fig:2DEGdispersion}
\end{figure}

{\it Beyond field-like torque in 2D.---}
We begin by identifying which torque components are allowed by symmetries, beyond the standard FL contribution. We examine non-magnetic point group \(C_{\infty v}\), which represents the highest symmetry group that supports Rashba SOC, exhibiting in-plane axial symmetry and an infinite set of mirror planes perpendicular to the 2D plane, while lacking mirror symmetry parallel to it. Notably, $C_{\infty v}$ encompasses the minimal set of potential torques in 2D systems, as any additional torque can only be enabled by reducing the symmetry group. To determine the minimal set of torques, the non-equilibrium spin density is expressed as a function of the orientation of the magnetization $\hat{\boldsymbol{m}}$ characterized by a polar angle $\theta$, and an azimuthal angle $\varphi$ relative to the applied electric field $\boldsymbol{\mathcal{E}}$ (see Fig.~\ref{fig:2DEGdispersion}-(a), inset). We retain only the components compatible with the system's symmetries and expand them up to second order in terms of their angular dependence \cite{manchon_current-induced_2019, garcia_ovalle_spin-orbit_2023}. This procedure enables us to  separate the non-equilibrium spin density in two contributions, $\boldsymbol{S} = \boldsymbol{S}_\text{I} + \boldsymbol{S}_\text{II}$, with
\begin{equation}
    \boldsymbol{S}_{\rm I} = \chi_\text{FL}^{} (1 - \xi_{\rm FL}^{} \sin^2\theta) \boldsymbol{H}_\text{SOC} - \chi_\text{DL}^{} \hat{\boldsymbol{m}} \times \boldsymbol{H}_\text{SOC} \text{ ,}
\end{equation}
corresponding to the standard FL and DL terms, causing the magnetization to precess and align along the effective SOC field \(\boldsymbol{H}_{\rm SOC}\equiv \hat{\boldsymbol{z}} \times \boldsymbol{\mathcal{E}}\). The adimensional parameter $\xi_\text{FL}^{}$ represents a second order term that modulates the magnetization precession as it approaches the plane. The second contribution reads:
\begin{equation}
    \boldsymbol{S}_{\rm II} = - \mathcal{E}(\chi_{\parallel}^{} \cos\varphi + \chi_{\perp}^{} \cos\theta \sin\varphi) \sin\theta \,\hat{\boldsymbol{z}} \text{ ,} 
\end{equation}
where $\chi_\parallel^{}$ generates an anisotropic damping of the in-plane magnetization with respect to the out-of-plane component, while $\chi_\perp^{}$ competes with the aforementioned stabilization along $\boldsymbol{H}_\text{SOC}$, favoring an alignment parallel or antiparallel to the current depending on the direction of $m_z$. These torques could be ascertained by observing non-trivial angular dependence in magnetoresistive experiments. 

Note that this procedure \emph{does not constitute a perturbative expansion}. Therefore, these torques, as well as those of higher order, can contribute on equal footing, and their existence and strength will depend on the underlying symmetries and competing fields. To discern their relevance, we develop a dual approach. Using the Kubo quantum transport framework, we first determine the non-equilibrium spin density at Fermi level $\varepsilon_\text{F}^{}$ using the Kubo-Bastin formula \cite{kubo_statistichal-mechanical_1957,bastin_quantum_1971},
\begin{equation}
    \boldsymbol{S}(\varepsilon_\text{F}^{}) = -2\hbar \int \text{d}\varepsilon f(\varepsilon) \text{Im} \,\text{Tr} \left[ \delta(\varepsilon - \hat{H}) \hat{\boldsymbol{s}} \,\partial_\varepsilon G^{+} (\hat{\boldsymbol{j}} \cdot \boldsymbol{\mathcal{E}}) \right] \text{ ,}
    \label{eq:S_KB}
\end{equation}
where $\hat{H}, \hat{\boldsymbol{s}}$ and $\hat{\boldsymbol{j}}$ are the Hamiltonian, spin and current density operators respectively, $f$ is the Fermi-Dirac distribution, and $G^+ = \text{lim}_{\eta \rightarrow 0} [\hat{H} - \varepsilon + \text{i}\eta]^{-1}$ is the retarded Green's function. Moreover, we distinguish Fermi-sea and Fermi-surface contributions adopting the decomposition proposed in Ref.~\cite{bonbien_symmetrized_2020}. We numerically compute Eq.~\eqref{eq:S_KB} employing a kernel polynomial method (KPM) expansion which includes the choice of a finite broadening $\eta$ \cite{weise_kernel_2006}. We simulate disordered systems via real-space linear scaling numerical methods, reaching a precision of $\eta \approx 15 \,\text{meV}$ in a system with $> 10^6$ orbitals \cite{garcia_real-space_2015, fan_linear_2021, supmat}. For pristine systems, we develop a $k$-space KPM calculation of Eq.~\eqref{eq:S_KB}, reaching $\eta \sim 1 ~\text{meV}$ precision in systems as large as $> 10^9$ orbitals \cite{supmat}. To compute $\boldsymbol{S}$ as a function of the magnetization, we select an optimal set of magnetization configurations, each requiring a separate Kubo calculation, resulting in 14 magnetization configurations for each system \cite{supmat}.

Simultaneously, we develop a semi-classical approach based on Boltzmann transport theory to understand the underlying microscopic mechanisms of these torques \cite{sinitsyn_semiclassical_2007}. Within this theory, the non-equilibrium spin density is
\begin{equation}
    \boldsymbol{S}(\varepsilon_\text{F}^{}) = \sum_\mu \int \frac{\text{d}^2\boldsymbol{k}}{(2\pi)^2} \Big[ \delta f_{\mu,\boldsymbol{k}} \boldsymbol{s}_{\mu,\boldsymbol{k}} + f(E_{\mu,\boldsymbol{k}}) \delta\boldsymbol{s}_{\mu,\boldsymbol{k}} \Big] \text{ ,}
    \label{eq:S_Boltzmann}
\end{equation}
with $E_{\mu,\boldsymbol{k}}$ and $| E_{\mu,\boldsymbol{k}} \rangle$ the eigenvalue and eigenstate respectively of an electron with crystal momentum $\boldsymbol{k}$ and band index $\mu$, and $\boldsymbol{s}_{\mu,\boldsymbol{k}} \equiv \langle E_{\mu,\boldsymbol{k}} | \hat{\boldsymbol{s}} | E_{\mu,\boldsymbol{k}} \rangle$ the mean value of the spin operator $\hat{\boldsymbol{s}}$, commonly referred to as the spin texture. The first term in Eq.~\eqref{eq:S_Boltzmann} represents the standard Boltzmann transport contribution stemming from the current-induced variation of the carrier occupation $\delta f_{\mu,\boldsymbol{k}}$. The origin of the FL torque has been extensively studied and arises from the current-induced drift of the Fermi surface, which combined with the helical spin texture of Rashba systems produces an in-plane spin density perpendicular to the current \cite{edelstein_spin_1990, manchon_theory_2008}, as illustrated in Fig.~\ref{fig:2DEGdispersion}-(b).

The second term in Eq.~\eqref{eq:S_Boltzmann} remains less understood. It originates from the adiabatic transport of Bloch states under an electric field and serves as a quantum mechanical correction to the semi-classical result \cite{xiao_berry_2010}. We here offer an intuitive interpretation of this aspect: as the carriers accelerate in the momentum-dependent Rashba field, a dynamic magnetic field develops within their rest frame. This field acts as a driving force that nudges the spin away from its equilibrium, producing a current-induced spin texture throughout the entire Fermi sea, as depicted in Fig.~\ref{fig:2DEGdispersion}-(c).

To illustrate this point, let us consider a Bloch Hamiltonian that incorporates a spinless component $H_{0,\boldsymbol{k}}$, together with an effective magnetic field \(\boldsymbol{B}_{\boldsymbol{k}}\), which encapsulates both the exchange and SOC fields. The Hamiltonian reads \(\hat{H}_{\boldsymbol{k}} = H_{0,\boldsymbol{k}} - \frac{1}{2} \boldsymbol{B}_{\boldsymbol{k}} \cdot \hat{\boldsymbol{s}}\) (we take the gyromagnetic ratio equals to unity). In equilibrium, the spin states align along $\boldsymbol{B}_{\boldsymbol{k}}$. When applying an electric field $\boldsymbol{\mathcal{E}}$, the electrons accelerate according to $\hbar\frac{\text{d}\boldsymbol{k}}{\text{d}t} = -e \boldsymbol{\mathcal{E}}$, with $e$ the elementary charge. The spin dynamics emerging from this non-equilibrium state is described by the Ehrenfest theorem, yielding $\frac{\text{d}}{\text{d}t} \boldsymbol{s}_{\mu,\boldsymbol{k}} + \hbar^{-1} \boldsymbol{B}_{\boldsymbol{k}} \times \boldsymbol{s}_{\mu,\boldsymbol{k}} = 0$. We decompose the spin texture into a component aligned with the instantaneous effective field and a perturbation induced by the current: $\boldsymbol{s}_{\mu,\boldsymbol{k}} = \mu \hat{\boldsymbol{B}}_{\boldsymbol{k}} + \delta \boldsymbol{s}_{\mu,\boldsymbol{k}}$, with $\mu=\pm$ the spin majority/minority band index. Within the linear response regime, the non-equilibrium spin texture reads
\begin{equation}
    \delta \boldsymbol{s}_{\mu,\boldsymbol{k}} = -\mu\frac{e\hbar}{B_{\boldsymbol{k}}^3} \boldsymbol{B}_{\boldsymbol{k}} \times \left( \boldsymbol{\mathcal{E}} \cdot \nabla_{\boldsymbol{k}} \right) \boldsymbol{B}_{\boldsymbol{k}} \text{ .}
    \label{eq:currentS}
\end{equation}
This result reveals a crucial point: the interplay between the equilibrium and dynamic magnetic fields results in a current-induced shift of the spin texture. Notably, this shift is the main contributor to the DL torque observed in 2D magnetic Rashba systems.

{\it Semi-classical torque mechanisms in a 2D electron gas.---} 
To highlight our theory's capabilities, we begin by showcasing the presence of a DL torque in an s-wave 2D electron gas (2DEG) with $C_{\infty v}$ symmetry. The magnetization is characterized by an exchange splitting $J_\text{ex}$ along the magnetization direction $\hat{\boldsymbol{m}}$, while the Rashba SOC field is helical with an isotropic amplitude $\Lambda_{\text{R},k}$. The Hamiltonian is $\hat{H}_{\boldsymbol{k}} = H_{0,k} - \frac{1}{2} \Lambda_{\text{R},k} \hat{\boldsymbol{\varphi}} \cdot \hat{\boldsymbol{s}} - \frac{1}{2} J_{\text{ex}} \hat{\boldsymbol{m}} \cdot \hat{\boldsymbol{s}}$,
with $H_{0,k}$ the kinetic term, and $\hat{\boldsymbol{\varphi}} = \hat{\boldsymbol{z}} \times \hat{\boldsymbol{k}}$. We determine the non-equilibrium spin density employing Boltzmann transport assuming an isotropic momentum relaxation time $\tau$ in the weak disorder regime. 
A perturbative expansion in terms of the magnetization direction is appropriate only if the effective magnetic field is dominated either by the exchange or Rashba term. We begin by defining the leading order effective magnetic field, $B_{0,k} = (J_\text{ex}^2 + \Lambda_{\text{R},k}^2)^{1/2}$, isotropic in momentum space. For a dominant exchange splitting, which is usually the experimental condition, $B_0 \approx J_\text{ex}$ is momentum independent, while for a dominant Rashba splitting $B_{0,k} \approx \Lambda_{\text{R},k}$. 

The conventional torques, represented in $\boldsymbol{S}_\text{I}$, show a similar behavior in both regimes. The FL torque is essentially determined by the spin helicity, $\boldsymbol{s}_{\mu,\boldsymbol{k}} \cdot \hat{\boldsymbol{\varphi}} \approx \mu \Lambda_{\text{R},k} / B_{0,k}$. It derives from the Fermi-surface contribution to Eq.~\eqref{eq:S_Boltzmann}, and is proportional to the density of states and the electron mobility following: 
\begin{equation}
    \chi_\text{FL}^{} = \mu \frac{e \tau k_\mu}{4\pi\hbar} \frac{\Lambda_{\text{R},k_\mu}}{B_{0,k_\mu}} \text{ ,} 
\end{equation}
where $\mu=\pm$ is the band index, the electron mobility is contained within $\tau$, and $k_\mu$ is the isotropic part of the Fermi momentum, which encodes the density of states, defined such that $\varepsilon_\text{F}^{} = H_{0,k_\mu} - \frac{\mu}{2} B_{0,k_\mu}$.

The DL torque on the other hand, is determined by the Fermi-sea integral of the current-induced spin texture $\delta \boldsymbol{s}_{\mu,\boldsymbol{k}}$. $\delta \boldsymbol{s}_{\mu,\boldsymbol{k}}$ emerges from the interaction between the equilibrium spin texture, and the variation of the effective magnetic field in the accelerating electron rest frame via spin-momentum locking, as given by Eq.~\eqref{eq:currentS}. Only the Rashba field $\Lambda_{\text{R}, k} \hat{\boldsymbol{\varphi}}$ contributes to the latter, yielding $\int \text{d} \varphi (\boldsymbol{\mathcal{E}} \cdot \nabla_{\boldsymbol{k}}) \boldsymbol{B}_{\boldsymbol{k}} = \frac{\pi}{k} \partial_k (k \Lambda_{\text{R},k})  \boldsymbol{H}_\text{SOC}$, whose interaction with the spin texture component parallel to the magnetization $\hat{\boldsymbol{m}}$  generates the DL torque
\begin{equation}
    \chi_\text{DL}^{} = \mu \frac{e J_\text{ex}}{4\pi} \int_0^{k_\mu} \text{d}k\, \frac{\partial_k (k \Lambda_{\text{R},k})}{B_{0,k}^3} \text{ .}
\end{equation}
This expression reveals that the DL torque is generated by the interplay between spin-momentum locking and exchange splitting throughout the entire Fermi sea. In the particular case of a dominant exchange splitting the effective field is momentum-independent, and the DL torque becomes $\chi_\text{DL}^{} = \frac{\hbar}{\tau J_\text{ex}} \chi_\text{FL}^{}$, thus offering a way to determine the momentum relaxation time via the $\chi_\text{FL}^{} / \chi_\text{DL}^{}$ ratio. Additionally, the interaction between the dynamic magnetic field and the equilibrium Rashba field produces an out-of-plane current-induced spin density inducing an unconventional torque $\chi_\parallel^{}$, which is only non-zero in the Rashba dominated regime \cite{supmat}.

We finally find that the second order torques are vanishing in parabolic systems, $\chi_\perp^{} = \xi_\text{FL}^{} = 0$. In the exchange dominated regime, the spin texture presents a dominant component aligned with $\hat{\boldsymbol{m}}$, while Rashba SOC introduces a helical component as well as an anisotropic modulation to the dominating one. Hence, the drift of the Fermi surface, in addition to the FL torque, also yields a non-equilibrium spin density parallel to the magnetization, which stems from the spin texture component parallel to $\hat{\boldsymbol{m}}$ and cannot exert any torque, but could be relevant for charge-to-spin conversion \cite{supmat}. Furthermore, our theory fully matches quantum simulations for the description of DL torque in a s-wave 2DEG \cite{supmat}.  

\begin{figure*}[t!]
    \centering
    \includegraphics[width=\textwidth]{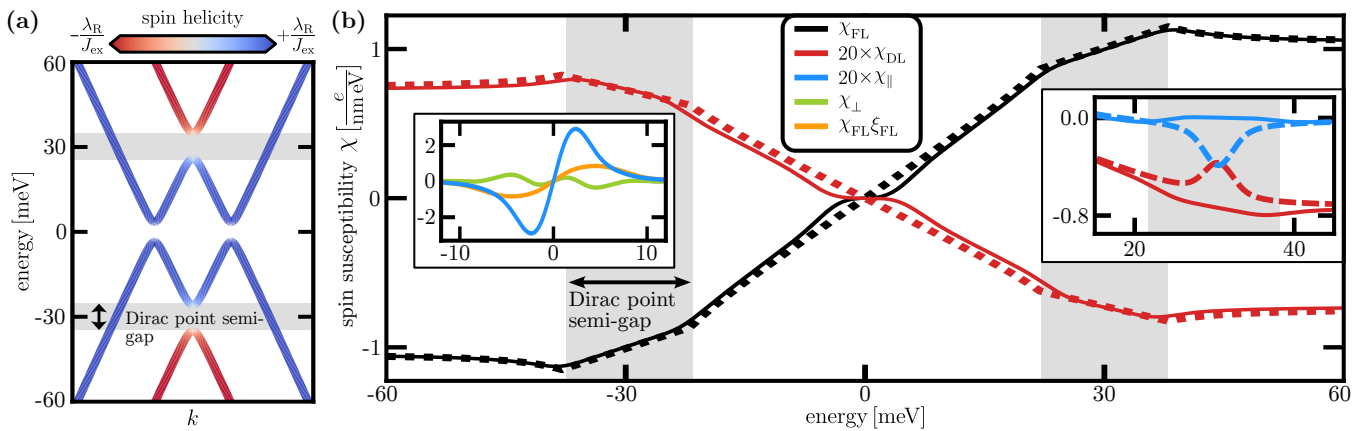}
    \caption{SOT in Dirac matter. \textbf{(a)} Band structure of Dirac system with Rashba SOC and exchange splitting [$\theta=0$]. \textbf{(b)} The conventional SOTs, $\chi_\text{FL}^{}$ and $\chi_\text{DL}^{}$, are driven by semi-classical effects, evinced by the agreement between semi-classical (dotted curves) and Kubo-Bastin (solid curves) frameworks. The insets show torque responses, computed by the Kubo framework, which are not captured by semi-classical effects. Left inset: All unvconventional torques show a strong enhancement near the band inversion at charge neutrality. Right inset: $\chi_\text{DL}^{}$ and $\chi_\parallel^{}$ shift from semi-classical to quantum-driven mechanisms at the Dirac point, as the semi-gap is dominated by the effective mass ($\Delta \neq 0$, solid curves) or Rashba SOC ($\Delta=0$, dashed curves) respectively. [$v \sim 10^6 m/s$, $\lambda_\text{R}^{} = 8 \,\text{meV}$, $J_\text{ex}^{} = 60 \,\text{meV}$, $\Delta = 8 \,\text{meV}$, $\eta = 2 \,\text{meV}$ at the band center and $\tau = \hbar / \eta$, except if indicated otherwise]}
    \label{fig:SOTgraphene}
\end{figure*}

{\it Unconventional torques in Dirac Matter.---}
We now use our theory to reveal unconventional torques in Dirac systems enabled by the additional angular momentum provided by the pseudospin degree of freedom. As a minimal model, we consider a $C_{\infty v}$ Dirac Hamiltonian doted with exchange splitting, given by $\hat{H}_{\boldsymbol{k}} = \hbar v \, \boldsymbol{k} \cdot \hat{\boldsymbol{\sigma}} + \Delta \sigma_z - \frac{1}{2} \lambda_\text{R} (\hat{\boldsymbol{s}} \times \hat{\boldsymbol{\sigma}}) \cdot \hat{\boldsymbol{z}} - \frac{1}{2} J_\text{ex} \hat{\boldsymbol{m}} \cdot \hat{\boldsymbol{s}}$, with $v$ the velocity of massless Dirac electrons, $\Delta$ their effective mass, and $\lambda_\text{R}^{}$ the Rashba SOC parameter, while the pseudospin is represented by Pauli vector $\hat{\boldsymbol{\sigma}}$. Such system may be realized by an insulating ferromagnet/graphene/TMD trilayer \cite{zollner_scattering-induced_2020}, where proximity effects induce SOC and exchange splitting \cite{konschuh_tight-binding_2010, gmitra_graphene_2015, leutenantsmeyer_proximity_2016}. Rashba splitting in graphene/TMD interfaces is typically of order $\sim 1 \,\text{meV}$ \cite{avsar_spinorbit_2014, wang_strong_2015, li_twist-angle_2019, naimer_twist-angle_2021}, while proximity with a ferromagnet can reach exchange splittings as large as $\sim 100 \,\text{meV}$ \cite{hallal_tailoring_2017, ibrahim_unveiling_2019}. We thus focus on the regime with a dominant exchange splitting.

Spin-momentum locking in Dirac electrons is mediated by the pseudospin. The 2DEG paradigm may be recovered in absence of spin-pseudospin correlations, i.e. $\langle \sigma_i s_j \rangle = \langle \sigma_i \rangle \langle s_j \rangle$, allowing us to define an effective Rashba field for each pseudospin polarized set of bands. This is the case far from band crossings, where Rashba SOC perturbatively imprints a helical component to the spin texture. Drastically different is the situation near band crossings, where Rashba SOC acts non-perturbatively by hybridizing bands of opposite pseudospin polarizations, while also inducing non-negligible spin-pseudospin correlations. These two regimes must be analyzed separately.

In the former regime, the pair of bands with opposite pseudospin polarization, or equivalently, opposite velocities, are decoupled. We thus recover the 2DEG paradigm by adequately adjusting for the corresponding kinetic Hamiltonian and Rashba field, allowing us to understand the FL and DL torques from simple band structure properties, shown in Fig.~\ref{fig:SOTgraphene}-(a): At energies larger than the magnetic splitting the spectrum presents two Fermi contours with positive velocity which compete due to their opposite helicities. Such competition is energy independent due to the linear dispersion. For energies lower than the magnetic splitting, the two Fermi contours present the same helicity, but with opposite velocity. The inner Fermi contour vanishes when approaching the spin-split Dirac point, where $\Delta$ opens a semi-gap. Within the semi-gapped energy window a single spin-helical band remains, representing an optimal condition for maximizing the SOT efficiency. The semi-classical results shown by dotted curves in Fig.~\ref{fig:SOTgraphene}-(b), which are in very good agreement with Kubo-Bastin calculations (shown in solid lines). Additionally, we recover the previously obtained relation for the momentum relaxation time and the torque ratio, $\tau= (\hbar/J_{\rm ex}) \chi_\text{FL}^{} / \chi_\text{DL}^{}$.

Near the band  inversion at charge neutrality the hybridization between bands of opposite velocity breaks the 2DEG paradigm. To elucidate the torques in this regime, the full quantum response provided by the Kubo-Bastin formula is required. The results are shown in Fig.~\ref{fig:SOTgraphene}-(b) left inset, where strong unconventional torques are seen near the Rashba gap, vanishing in the 2DEG case. The origin of these torques is two-fold. Despite the dominant exchange splitting, Rashba SOC acts non-perturbatively and induces strong anisotropies in the dispersion. Furthermore, non-negligible spin-pseudospin correlations quench the spin texture \cite{Tuan2014,de_moraes_emergence_2020} and modify the coupled spin-pseudospin dynamics \cite{supmat}. These two features are only possible due to the additional pseudospin degree of freedom.

At the Dirac point, the SOT physics may shift from semi-classical to quantum driven effects. While the effective mass $\Delta$ acts separately on each spin-split Dirac cone, Rashba SOC couples them favoring spin-pseudospin entanglement \cite{Tuan2014,de_moraes_emergence_2020}. At the spin-split Dirac point the spectrum is semi-gapped, thus transport should be dominated by the band of the opposite cone, whose physics is mainly determined by the kinetic term of the Hamiltonian, insensitive to $\Delta$. This is indeed the case for massive Dirac electrons, as shown in Fig.~\ref{fig:SOTgraphene}-(b) right inset (solid curves), where $|\chi_\text{DL}^{}|$ increases across the semi-gap while $\chi_\parallel^{}$ remains negligible. However, the torques change dramatically for massless electrons, as $|\chi_\text{DL}^{}|$ is minimal and $|\chi_\parallel^{}|$ peaks within the semi-gap, shown in Fig.~\ref{fig:SOTgraphene}-(b) right inset (dashed curves). This qualitative change is generated due to spin-pseudospin entanglement dominating the gap, indicating a large torque originated from a vanishing Fermi contour.

\begin{figure}
    \centering
    \includegraphics[width=\columnwidth]{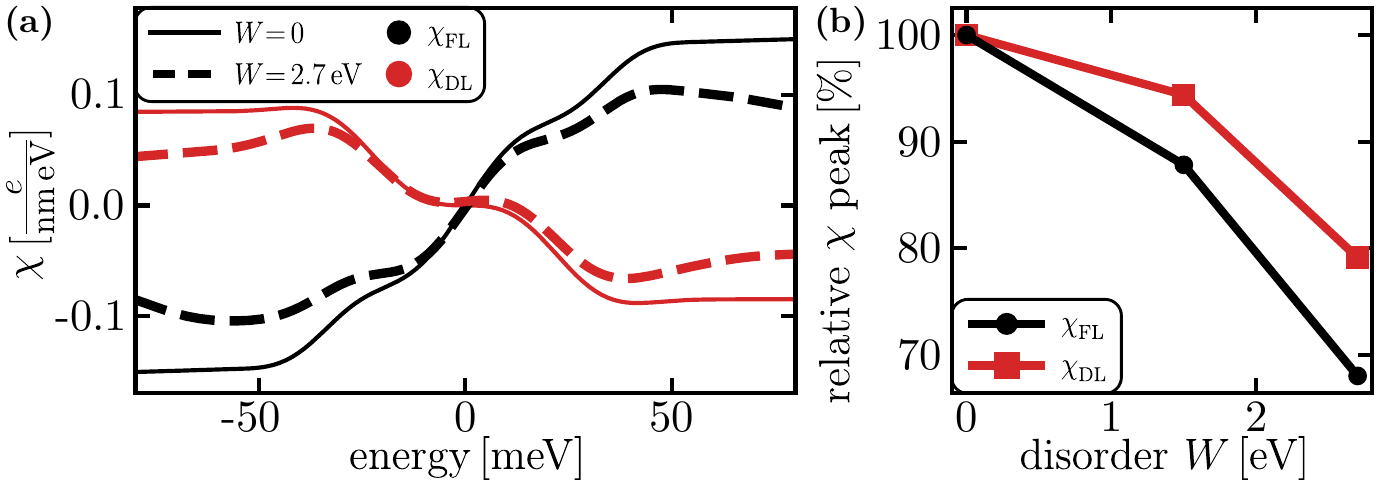}
    \caption{The FL and DL torques are robust against Anderson disorder (of strength $W$). \textbf{(a)} SOT for pristine (solid) and $W=2.7\,\text{eV}$ disordered (dashed) systems. \textbf{(b)} Peak torque values relative to the pristine system. [$v\sim10^{6} m/s$, $\lambda_\text{R}=10\,\text{meV}$, $J_\text{ex}=40\,\text{meV}$, $\Delta=5\,\text{meV}$ and $\eta = 15\,\text{meV}$ at the band center.]}
    \label{fig:SOTdisorder}
\end{figure}

We finally compute the Kubo-Bastin response via linear scaling numerical methods \cite{fan_linear_2021, garcia_real-space_2015, supmat} considering real-space Anderson disorder of strength $W$. The results are shown in Fig.~\ref{fig:SOTdisorder}, focusing on the FL and DL torques. For $W$ as high as $2.7 \,\text{eV}$, the $\chi_\text{FL}^{}$ and $\chi_\text{DL}^{}$ peaks are only reduced to 68\% and 79\% of their respective pristine values. Dictated by a Fermi-sea contribution, $\chi_\text{DL}^{}$ shows a better resilience than $\chi_\text{FL}^{}$.

In conclusion, we have developed a general theory of SOT for interfaces involving low dimensional (Dirac) materials with broken inversion symmetry. Novel types of SOT components, resilient to disorder effects, have been found to emerge and superimpose, enhancing the torque capability of these interfaces for more efficient magnetization switching, enabling further exploration of ultralow energy memory and spintronic applications in ultracompact devices.

\begin{acknowledgments}
The authors thank D. García Ovalle and A. Manchon for fruitful discussions. 
The authors acknowledge funding from Ministerio de Ciencia e Innovacion (MCIN) under grant PID2022-138283NB-I00/MCIN/AEI/10.13039/501100011033 and the European Regional Development Fund.
J.M.D. acknowledges support from MCIN (grant FPI PRE2021-097031). 
J.H.G. acknowledge funding from the European Union (ERC, AI4SPIN, 101078370). 
S.R and J.H.G, acknowledge grant PCI2021-122035-2A-2 funded by MCIN/AEI/10.13039/501100011033 and European Union ``NextGenerationEU/PRTR”, funding from the European Union’s Horizon 2020 research and innovation programme under grant No 881603, and the support from Departament de Recerca i Universitats de la Generalitat de Catalunya. 
ICN2 is funded by the CERCA Programme/Generalitat de Catalunya and supported by the Severo Ochoa Centres of Excellence programme, Grant CEX2021-001214-S, funded by MCIN/AEI/10.13039.501100011033.
\end{acknowledgments}

%

\clearpage
\onecolumngrid

\section{Supplemental Notes on Kubo calculations}

\noindent
We present additional details for the Kubo-Bastin calculations of SOT. In order to distill the macroscopic torques according to their dependence on the magnetization direction, we select an evenly spaced sample of $2^M + 1$ $\theta$ points and $2M$ $\varphi$ points, with $M\in\mathbb{N}$ the order of the $\boldsymbol{S}$ expansion.

Calculations of disordered systems, as presented in Fig.~3 of the main text, are performed by real-space simulations via linear scaling numerical methods~\cite{fan_linear_2021, garcia_real-space_2015}. We consider a finite system of $1024\times1024$ unit cells including random real-space disorder, and use a stochastic approximation of the trace for each disorder configuration. We compute the non-equilibrium spin density for a band center broadening of $15 \,\text{meV}$, corresponding to $1867$ Chebyshev moments, and average over $11$ disorder configurations.

In order to reach higher energy resolution, we develop a $k$-space representation of the KPM expansion of the Kubo-Bastin formula for pristine systems, as presented in Fig.~2 of the main text. We consider a finite system of $14691 \times 14691$ unit cells, which in the pristine limit corresponds to a $14691 \times 14691$ $k$-point grid sampling the Brillouin zone. Aiming to describe the response of the system near the Dirac points, we keep only the $\boldsymbol{k}$ points with eigenvalues within a $\pm 300 \,\text{meV}$ energy range from charge neutrality, reducing the computational cost in more than $99\%$. Furthermore, focusing on a finite energy window about charge neutrality allows us to reach lower broadening with a smaller Chebyshev expansion. We compute the non-equilibrium spin density for a band center broadening of $2 \,\text{meV}$, corresponding to $581$ Chebyshev moments.

\section{Semi-classical calculations of SOT in a 2D electron gas}

\noindent
In this section we present the complete calculations for the semi-classical spin density in a 2D electron gas. We analyze as well the specific case of the 2D electron gas expanded to lowest order in momentum, including a comparison between both semi-classical and Kubo-Bastin methods.

We begin expanding the relevant physical magnitudes in series in terms of the magnetization direction. Note that the expansion must be physically sustained by a weak interplay between the exchange and Rashba fields, which we incorporate at the end of the calculation. We define $B_{0,k} = (J_\text{ex}^2 + \Lambda_{\text{R},k}^2)^{1/2}$ the leading order term of the magnetic field. The dispersion of band $\mu$ is 
\begin{equation}
    E_{\mu,\boldsymbol{k}} = H_{0,k} - \frac{\mu}{2}B_{0,k} - \frac{\mu}{2} \frac{J_\text{ex} \Lambda_{\text{R},k}}{B_{0,k}} (\hat{\boldsymbol{m}} \cdot \hat{\boldsymbol{\varphi}}) \text{ .}
\end{equation}
For the Fermi momentum of band $\mu$ we assume an ansatz of the form $k_{\text{F},\mu}(\varphi) = k_\mu + \delta k_\mu (\hat{\boldsymbol{m}} \cdot \hat{\boldsymbol{\varphi}})$, with $k_\mu$ the leading contribution isotropic in $k$-space and independent of the magnetization direction. Intersection with Fermi level $\varepsilon_\text{F}^{}$ yields the dominant contribution $k_\mu$ such that $\varepsilon_\text{F}^{} = H_{0,k_\mu} - \frac{\mu}{2} B_{0,k_\mu}$, and the first order correction
\begin{equation}
    \delta k_\mu = \frac{\mu}{2}\frac{J_\text{ex} \Lambda_{\text{R},k_\mu}}{B_{0,k_\mu} \partial_k (H_{0,k_\mu} - \frac{\mu}{2} B_{0,k_\mu})} \text{ .}
\end{equation}
Caution must be taken near the band edges, as if the band edge occurs at a finite momentum the Fermi surface is strongly anisotropic, thus escaping the validity regime of the present calculations.

We calculate the non-equilibrium spin density within the Boltzmann semi-classical framework, as given by Eq.~(4) of the main text. At zero temperature $f_{\mu,\boldsymbol{k}} = \Theta(\varepsilon_\text{F}^{} - E_{\mu,\boldsymbol{k}})$, with $\Theta$ the Heavyside step function. Assuming a momentum relaxation time $\tau$, the current-induced variation of the carrier occupation is $\delta f_{\mu,\boldsymbol{k}} = \frac{e\tau}{\hbar} \delta(\varepsilon_\text{F}^{} - E_{\mu,\boldsymbol{k}}) (\boldsymbol{\mathcal{E}} \cdot \nabla_{\boldsymbol{k}}) E_{\mu,\boldsymbol{k}}$. The Fermi-surface and Fermi-sea contributions to the non-equilibrium spin density respectively are
\begin{subequations}
\begin{equation}
\begin{split}
    \boldsymbol{S}_{\mu}^\text{surf} =&  \frac{e\tau}{4\pi^2\hbar} \int_0^{2\pi} \text{d}\varphi \,k \frac{(\boldsymbol{\mathcal{E}} \cdot \nabla_{\boldsymbol{k}}) E_{\mu,k}}{|\nabla_{\boldsymbol{k}} E_{\mu,k}|} \boldsymbol{s}_{\mu,\boldsymbol{k}} \Bigg|_{k=k_\text{F},\mu} \\
    =& \mu\frac{e\tau k_\mu}{4\pi\hbar} \frac{\Lambda_{\text{R},k_\mu}}{B_{0,k_\mu}} \left( \hat{\boldsymbol{z}} \times \boldsymbol{\mathcal{E}} - \frac{J_\text{ex}^2}{B_{0,k_\mu}^2} ( \hat{\boldsymbol{z}} \times \boldsymbol{\mathcal{E}} \cdot \hat{\boldsymbol{m}}) \,\hat{\boldsymbol{m}} \right) \text{ ,}
\end{split}
\label{eq:Ssurf_supmat}
\end{equation}
\begin{equation}
\begin{split}
    \boldsymbol{S}_\mu^\text{sea} =& \frac{1}{4\pi^2} \int_0^{2\pi} \text{d}\varphi \int \text{d}k\, \left[ \Theta(k_\mu - k) + \delta(k_\mu - k) \delta k_\mu (\hat{\boldsymbol{m}} \cdot \hat{\boldsymbol{\varphi}}) \right] k\, \delta\boldsymbol{s}_{\mu,\boldsymbol{k}} \\
    =& -\mu \frac{J_\text{ex}}{4\pi} \left[ \hat{\boldsymbol{m}} \times (\hat{\boldsymbol{z}} \times \boldsymbol{\mathcal{E}}) \int_0^{k_\mu} \text{d}k\, \frac{\partial_k (k \Lambda_{\text{R},k})}{B_{0,k}^3} + (\hat{\boldsymbol{m}} \cdot \boldsymbol{\mathcal{E}}) \hat{\boldsymbol{z}} \left( \frac{\Lambda_{\text{R},k_\mu}^3}{B_{0,k_\mu}^4 \partial_k(2\mu H_{0,k_\mu} - B_{0,k_\mu})} - \int_0^{k_\mu} \text{d}k\, \frac{3 \Lambda_{\text{R},k}^3}{B_{0,k}^5} \right) \right] \text{ .}
\end{split}
\label{eq:Ssea_supmat}
\end{equation}
\end{subequations}
The first term in Eqs. \eqref{eq:Ssurf_supmat} and \eqref{eq:Ssea_supmat} yield the final results for the field-like and damping-like torques, presented in Eqs. (6) and (7) of the main text respectively. The unconventional torque $\chi_\parallel^{}$ stems from the second term in Eq.~\eqref{eq:Ssea_supmat}, which is only non-negligible in the Rashba-dominated regime. Finally, we note that the Fermi-Surface contribution yields a spin density of the form $\propto (\hat{\boldsymbol{z}} \times \boldsymbol{\mathcal{E}} \cdot \hat{\boldsymbol{m}}) \, \hat{\boldsymbol{m}}$; which, though is allowed by symmetries, does not represent a SOT as it is parallel to the magnetization, and is thus excluded from Eqs.~(1) and (2) of the main text.

The semi-classical non-equilibrium spin density is calculated in a weak disorder limit as $\tau \rightarrow \infty$. In the Kubo-Bastin formalism this regime is represented by a vanishing broadening $\eta \rightarrow 0$. Under these conditions, the Fermi surface contribution to the Kubo-Bastin formula, as indicated by the Bonbien \& Manchon decomposition \cite{bonbien_symmetrized_2020}, reduces to that obtained in Boltzmann transport. On the other hand, the Kubo-Bastin Fermi sea contribution reads
\begin{equation}
\begin{split}
    \boldsymbol{S}^\text{sea} \rightarrow& -\hbar \sum_{\mu,\mu'} \int \frac{\text{d}^2 \boldsymbol{k}}{(2\pi)^2} \frac{f(E_{\mu,\boldsymbol{k}}) - f(E_{\mu',\boldsymbol{k}})}{(E_{\mu,\boldsymbol{k}} - E_{\mu',\boldsymbol{k}})^2} \,\text{Im}\, \langle E_{\mu,\boldsymbol{k}} | \hat{\boldsymbol{j}} \cdot \boldsymbol{\mathcal{E}} | E_{\mu',\boldsymbol{k}} \rangle \langle E_{\mu',\boldsymbol{k}} | \hat{\boldsymbol{s}} | E_{\mu,\boldsymbol{k}} \rangle \\ 
    =& -\hbar \sum_{\mu=\pm} \int \frac{\text{d}^2\boldsymbol{k}}{(2\pi)^2} \frac{2f(E_{\mu,\boldsymbol{k}})}{B_{\boldsymbol{k}}^2} \,\text{Im}\, \langle E_{\mu,\boldsymbol{k}} | \hat{\boldsymbol{j}} \cdot \boldsymbol{\mathcal{E}} | E_{-\mu,\boldsymbol{k}} \rangle \langle E_{-\mu,\boldsymbol{k}} | \hat{\boldsymbol{s}} | E_{\mu,\boldsymbol{k}} \rangle \\
    =& \sum_{\mu=\pm} \int \frac{\text{d}^2\boldsymbol{k}}{(2\pi)^2} f(E_{\mu,\boldsymbol{k}} ) \delta \boldsymbol{s}_{\mu,\boldsymbol{k}}
\end{split}
\end{equation}
A straight-forward calculation reveals that $\text{Im}\, \langle E_{\mu,\boldsymbol{k}} | \hat{\boldsymbol{j}} \cdot \boldsymbol{\mathcal{E}} | E_{-\mu,\boldsymbol{k}} \rangle \langle E_{-\mu,\boldsymbol{k}} | \hat{\boldsymbol{s}} | E_{\mu,\boldsymbol{k}} \rangle = \frac{e}{2} \boldsymbol{s}_{\boldsymbol{k}} \times (\boldsymbol{\mathcal{E}} \cdot \nabla_{\boldsymbol{k}}) \boldsymbol{B}_{\boldsymbol{k}}$, proportional to the current-induced spin texture, presented in Eq.~(5) of the main text. Thus, the Fermi sea contribution in the Kubo-Bastin formalism corresponds exactly to that of the Boltzmann framework.

\subsection{2DEG: Parabolic dispersion}

We analyze the particular case of the 2DEG hamiltonian with a parabolic kinetic term, characterized by the effective mass $m_\text{eff}$, and a linear Rashba field $\Lambda_{\text{R},k} = \lambda_\text{R} k$. The Hamiltonian is
\begin{equation}
    \hat{H}_{\boldsymbol{k}} = \frac{k^2}{2 m_\text{eff}} - \frac{\lambda_\text{R}^{} k}{2} \hat{\boldsymbol{\varphi}} \cdot \hat{\boldsymbol{s}} - \frac{J_\text{ex}^{}}{2} \hat{\boldsymbol{m}} \cdot \hat{\boldsymbol{s}} \text{ ,}
\end{equation}
where we use $\hbar=1$. The isotropic part of the Fermi momentum of band $\mu$, at Fermi level $\varepsilon_\text{F}^{}$ is
\begin{equation}
    k_\mu^2 = 2 m_\text{eff} \varepsilon_\text{F}^{} + \frac{m_\text{eff}^2 \lambda_\text{R}^2}{2} + \mu m_\text{eff} \left( J_\text{ex}^2 + 2 m_\text{eff} \lambda_\text{R}^2 \varepsilon_\text{F}^{} + \frac{m_\text{eff}^2 \lambda_\text{R}^4}{4} \right)^{1/2} \text{ .}
\end{equation}
The non-equilibrium spin density of band $\mu$ is given by
\begin{subequations}
\begin{equation}
    \chi_\text{FL}^{} = \mu \frac{e\tau \lambda_\text{R}^{} k_\mu^2}{4\pi\hbar (J_\text{ex}^2 + \lambda_\text{R}^2 k_\mu^2)^{1/2}} \text{ ,}
\end{equation}
\begin{equation}
    \chi_\text{DL}^{} = \mu \frac{e}{2\pi \lambda_\text{R}^{}} \left[ 1 - \frac{J_\text{ex}^{}}{(J_\text{ex}^2 + \lambda_\text{R}^2 k_\mu^2)^{1/2}} \right] \text{ ,}
\end{equation}
\begin{equation}
    \chi_\parallel^{} = -\mu \frac{e}{2\pi\lambda_\text{R}^{}} \left[ 1 - \frac{J_\text{ex}^{}}{(J_\text{ex}^2 + \lambda_\text{R}^2 k_\mu^2)^{1/2}} - \frac{J_\text{ex}^{} \lambda_\text{R}^2 k_\mu^2}{2 (J_\text{ex}^2 + \lambda_\text{R}^2 k_\mu^2)^{3/2}} \left( 1 + \frac{m_\text{eff} \lambda_\text{R}^2}{2\mu(J_\text{ex}^2 + \lambda_\text{R}^2 k_\mu^2)^{1/2} -  m_\text{eff} \lambda_\text{R}^2} \right) \right] \text{ ,}
\end{equation}
\begin{equation}
    \chi_\text{torque-less}^{} = \mu \frac{e\tau J_\text{ex}^2 \lambda_\text{R}^{} k_\mu^2}{4\pi\hbar (J_\text{ex}^2 + \lambda_\text{R}^2 k_\mu^2)^{3/2}} \text{ ,}
\end{equation}
\end{subequations}
with $\chi_\text{torque-less}^{}$ representing the spin density contribution $\boldsymbol{S}_\text{torque-less}^{} = - \chi_\text{torque-less}^{}(\hat{\boldsymbol{m}} \cdot \hat{\boldsymbol{z}} \times \boldsymbol{\mathcal{E}}) \, \hat{\boldsymbol{m}}$ which cannot exert torque, while $\chi_\perp^{} = \xi_\text{FL}^{} = 0$. 

\begin{figure}
    \centering
    \includegraphics[width=0.95\textwidth]{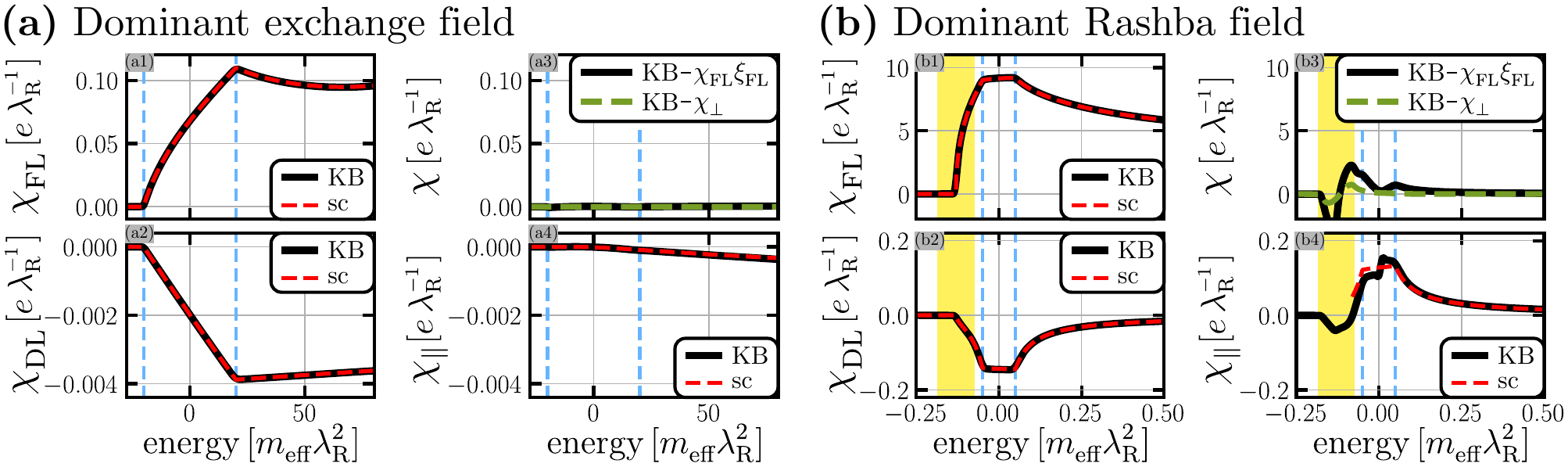}
    \caption{Non-equilibrium spin density in 2DEG, comparing semi-classical [sc] and Kubo-Bastin [KB] results. The Rashba pseudogap is delimited by $E_{\mu,k=0} = \frac{\mu}{2} J_\text{ex}^{}$ (dashed blue lines). \textbf{(a)} Dominant exchange field [$J_\text{ex}^{} = 40 m_\text{eff} \lambda_\text{R}^2$, and $1.5 m_\text{eff}\lambda_\text{R}^2$ Kubo broadening at the band center, corresponding to 187 Chebyshev moments]. \textbf{(b)} Dominant Rashba field [$J_\text{ex}^{} = 0.1 m_\text{eff} \lambda_\text{R}^2$, and $0.015 m_\text{eff} \lambda_\text{R}^2$ Kubo broadening at the band center, corresponding to 340 Chebyshev moments]. The yellow shaded region shows the energy range in which the Fermi momentum is not defined in all $\varphi$ directions. Panels (a3), (b3) only show Kubo-Bastin torques, as the corresponding semi-classical torques are zero. All calculations performed fixing $m_\text{eff}$, with the energy and distance scales determined by $m_\text{eff}\lambda_\text{R}^2$ and $(m_\text{eff} \lambda_\text{R})^{-1}$ respectively}
    \label{fig:SOT-2DEG}
\end{figure}

We compare the results obtained analytically via semi-classical calculations with those obtained from Kubo-Bastin simulations. For a dominant exchange field the results are shown in Fig.~\ref{fig:SOT-2DEG}-(a), revealing an excellent agreement between both methods. All non-zero contributions, except for $\chi_\parallel^{}$, are maximal at the high energy edge of the Rashba pseudogap as a single spin-helical band takes part in transport. $\chi_\parallel^{}$ on the other hand is supressed by the exchange field and increases at higher energies along with the Rashba field magnitude. These results show that all the SOT mechanisms are captured by the semi-classical theory.

The analysis is more complicated for a dominant Rashba field, whose results are shown in Fig.~\ref{fig:SOT-2DEG}-(b), which cannot dominate throughout the entire spectrum as it is proportional to the momentum. Furthermore, the dispersion shifts from Rashba to kinetic dominance yielding a band edge at a finite momentum. To leading order the band edge is $\varepsilon_\text{edge} = -\frac{1}{8} m_\text{eff} \lambda_\text{R}^2 - \frac{1}{2} m_\text{eff}^{-1} \lambda_\text{R}^{-2} J_\text{ex}^2$, which occurs at momentum $k_\text{edge} = \frac{1}{2} ( m_\text{eff}^2 \lambda_\text{R}^4 - 4 J_\text{ex}^2 )^{1/2}$ for band $\mu=+$. Within an energy window of width $J_\text{ex} \lambda_\text{R} k_\text{edge} (1 - m_z^2) ( J_\text{ex} + \lambda_\text{R}^2 k_\text{edge}^2 )^{-1/2}$ about $\varepsilon_\text{edge}$, highlighted in yellow in Fig.~\ref{fig:SOT-2DEG}-(b) the Fermi momentum is not defined for all $\varphi$ directions. Due to the breakdown of the approximations, we do not expect accurate results for the unconventional torques near $\varepsilon_\text{edge}$ as they involve a perturbative treatment of the dispersion anisotropies. Note that this limitation does not lie in the semi-classical framework itself, which was proven to yield equivalent results to the Kubo-Bastin framework, but lie in the approximations performed in order to isolate the separate contributions to the spin density. Fig.~\ref{fig:SOT-2DEG}-(b) reveals that the conventional FL and DL torques are fully captured by the semi-classical theory. The semi-classical results for the unconventional torques are valid outside the yellow shaded region, showing good agreement with Kubo-Bastin calculations, specially for energies higher than the Rashba pseudo-gap.

\section{Spin dynamics and semi-classical spin-orbit torque in Dirac system}

\noindent
In this section we present the calculations for the semi-classical torques in a Dirac system.

The spin-pseudospin dynamics are given by coupled equations, which not only involve the spin and pseudospin textures, but also spin-pseudospin correlations. The dynamic equations are
\begin{subequations}
\begin{equation}
    \frac{\text{d}}{\text{d}t} \langle \boldsymbol{\hat{\sigma}} \rangle - 2 (\hbar v \boldsymbol{k} + \Delta \hat{\boldsymbol{z}}) \times \langle \hat{\boldsymbol{\sigma}} \rangle + \lambda_\text{R}^{} \langle (\hat{\boldsymbol{z}} \times \hat{\boldsymbol{s}}) \times \hat{\boldsymbol{\sigma}} \rangle  = 0 \text{ ,}
\end{equation}
\begin{equation}
    \frac{\text{d}}{\text{d}t} \langle \hat{\boldsymbol{s}} \rangle + J_\text{ex}^{} \hat{\boldsymbol{m}} \times \langle \hat{\boldsymbol{s}} \rangle - \lambda_\text{R}^{} \langle (\hat{\boldsymbol{z}} \times \hat{\boldsymbol{\sigma}}) \times \hat{\boldsymbol{s}} \rangle = 0 \text{ .}
\end{equation}
\end{subequations}
Within a perturbative treatment of Rashba SOC, the spin and pseudospin textures are, to leading order, independently defined by the exchange and bare Dirac Hamiltonians respectively. The states present a strong spin and pseudospin polarization $|\boldsymbol{s}| = |\boldsymbol{\sigma}| \approx 1 + \mathcal{O}(\lambda_\text{R}^{})$, and the spin-pseudospin correlations are disentangled as $\lambda_\text{R}^{} \langle \hat{\sigma}_i \hat{s}_j \rangle = \lambda_\text{R}^{} \langle \hat{\sigma}_i \rangle \langle \hat{s}_j \rangle$. We may now define the effective magnetic field as a classical variable, namely, $\boldsymbol{B}_{\boldsymbol{k}} = \boldsymbol{B}_0 - \lambda_\text{R} \hat{\boldsymbol{z}} \times \boldsymbol{\sigma}_{\boldsymbol{k}}$, which presents a spin-pseudospin coupling term proportional to $\lambda_\text{R}$, and a SOC-independent term $\boldsymbol{B}_0 = J_\text{ex} \hat{\boldsymbol{m}}$. Equivalently, we define a pseudomagnetic field $\boldsymbol{\beta}_{\boldsymbol{k}} = \boldsymbol{\beta}_{0,\boldsymbol{k}} + \lambda_\text{R} \hat{\boldsymbol{z}} \times \boldsymbol{s}_{\boldsymbol{k}}$, with $\boldsymbol{\beta}_{0,\boldsymbol{k}} = -2(\hbar v \boldsymbol{k} + \Delta \hat{\boldsymbol{z}})$. Note that a separate magnetic field is defined for each pseudospin polarized set of bands. The spin and pseudospin textures are obtained solving the equations in equilibrium ($\frac{\text{d}}{\text{d}t} \boldsymbol{\sigma}_{\boldsymbol{k}} = \frac{\text{d}}{\text{d}t} \boldsymbol{s}_{\boldsymbol{k}} = 0$), yielding to first order
\begin{subequations}
\begin{equation}
    \boldsymbol{s}_{\mu \nu, \boldsymbol{k}} = \mu \left[ \hat{\boldsymbol{B}}_0 - \nu \frac{\lambda_\text{R}}{B_0} \hat{\boldsymbol{B}}_0 \times \left( \hat{\boldsymbol{B}}_0 \times \left( \hat{\boldsymbol{z}} \times \hat{\boldsymbol{\beta}}_{0,\boldsymbol{k}} \right) \right) \right] \text{ ,}
\end{equation}
\begin{equation}
    \boldsymbol{\sigma}_{\mu \nu, \boldsymbol{k}} = -\nu \left[ \hat{\boldsymbol{\beta}}_{0,\boldsymbol{k}} - \mu \frac{\lambda_\text{R}}{\beta_{0,\boldsymbol{k}}} \hat{\boldsymbol{\beta}}_{0,\boldsymbol{k}} \times \left( \hat{\boldsymbol{\beta}}_{0,\boldsymbol{k}} \times \left( \hat{\boldsymbol{z}} \times \hat{\boldsymbol{B}}_0 \right) \right) \right] \text{ ,}
\end{equation}
\end{subequations}
with $\nu, \mu = \pm$ the particle/hole and spin majority/minority band indexes respectively. The strong pseudospin polarization of the bands allows us to treat it as a classical variable. We rewrite the full Hamiltonian as a collection of two separate Hamiltonians, one for each pseudospin polarization, each with spin as its only internal degree of freedom, namely $\hat{H}_{\boldsymbol{k}} = \bigoplus_{\nu=\pm} \hat{H}_{\nu,\boldsymbol{k}}$. The pseudospin-polarized Hamiltonians $\hat{H}_{\nu,\boldsymbol{k}}$ present the same structure as the 2DEG Hamiltonian as given in Eq.~(7) of the main text, with $H_{0,\nu,\boldsymbol{k}} = \nu \sqrt{(\hbar vk)^2 + \Delta^2}$ and $\Lambda_{\text{R},\nu,\boldsymbol{k}} = -\nu \hbar v k \lambda_\text{R} [(\hbar vk)^2 + \Delta^2]^{-1/2}$. From this point the results obtained for the 2DEG case may be used, taking care of performing the sum over all pseudospin polarizations. Note that the performed approach, though consistent with a perturbative treatment of Rashba SOC to first order, does not accurately capture the behaviour of the system near band crossings where SOC acts non-perturbatively, as evinced by the lack of a gap at the band crossing between both spin-split Dirac cones.

\end{document}